# A Volume Clearing Algorithm for Muon Tomography

D. Mitra (IEEE Senior Member), K. Day, and M. Hohlmann (IEEE Member)

Florida Institute of Technology, Melbourne, FL 32901

**Abstract**: The primary objective is to enhance muon-tomographic image reconstruction capability by providing distinctive information in terms of deciding on the properties of regions or voxels within a probed volume '$V$' during any point of scanning: threat type, non-threat type, or not-sufficient data. An algorithm (*MTclear*) is being developed to ray-trace muon tracks and count how many straight tracks are passing through a voxel. If a voxel '$v$' has sufficient number of straight tracks ($t$), then '$v$' is a non-threat type voxel, unless there are sufficient number of scattering points ($p$) in '$v$' that will make it a threat-type voxel. The algorithm also keeps track of voxels for which not enough information is known: where $p$ and $v$ both fall below their respective threshold parameters. We present preliminary results showing how the algorithm works on data collected with a Muon Tomography station based on gas electron multipliers operated by our group. The MTclear algorithm provides more comprehensive information to a human operator or to a decision algorithm than that provided by conventional muon-tomographic reconstruction algorithms, in terms of qualitatively determining the threat possibility from a probed volume. This is quite important because only low numbers of cosmic ray source muons are typically available in nature for tomography, while a quick determination of threats is essential.

## 1.  Overview

Muon tomography stations (MTS) are being developed all over the world for non-invasive detection of high-atomic number (Z) materials [1,2]. One of the objectives of Homeland Security is to find nuclear contraband transported clandestinely across geopolitical borders. Muons are elementary particles with the same charges as electrons and positrons, but about two hundred times more massive. They are generated continuously with typical energies of several GeV in the upper atmosphere by cosmic rays. Muons scatter from nuclei of atoms, and the distribution of scattering angles depends on the atomic number of the respective nucleus. Muon detector arrays can find muon tracks with minimally affecting them. Consequently, the angle of a scattered muon can be determined precisely by deploying position-sensitive detector arrays with electronic coincidence circuitry surrounding a probed volume $V$, e.g. a cargo container. Over the last decade, many experimental MTS have been developed and tomographic algorithms were proposed for imaging the high-Z target objects [3,4,5,6]. The main challenge here is in dealing with the low number of available cosmic-ray generated muons (one muon per cm$^2$ per minute at sea level). This leads to a long wait time for fully imaging a probed volume.

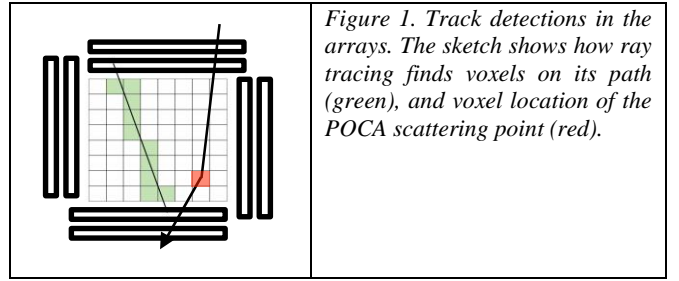

*Figure 1. Track detections in the arrays. The sketch shows how ray tracing finds voxels on its path (green), and voxel location of the POCA scattering point (red).*

In this work, we try to approach the tomographic problem with a reverse hypothesis: *a probed volume V may be cleared more efficiently during scanning when it does not contain any high-Z object than accurately imaging such objects when they are present*. We have conducted some preliminary investigation towards clearing the voxels (3D pixels) in $V$ while detecting the scattered events. We utilize non-scattered straight muon tracks that are normally ignored in a conventional algorithm meant for threat object detection.

## 2.  Device and Methods

Our group has developed and is operating a prototype MTS [4] with an array of eight 30×30 cm$^2$ Gas Electron multiplier (GEM) detectors surrounding $V$ (Fig. 1) [2].

A unique feature of the GEM micropattern gas detectors is that they have a much higher spatial resolution (~ 100 μm) compared to that (~ 300 μm) of more conventional detectors based on drift tubes, used in the first generation MTSs. Another additional feature of this station is that it deploys detector arrays on top, bottom, left and right sides of $V$ providing a better coverage. Fig. 1 shows the schematic diagram of the detector arrays and the muon tracking for both scattering and non-scattering events.

The MTS provides the 3D point of detection and the direction of a muon track on two sides of probed volume $V$ where a coincidence event is generated. Each of the $N$ events in the set $E$ is ($A_i$, $D_i$, $A_o$, $D_o$) where $A_i$, $D_i$, are the angle and the point on a detector array of the incoming muon, and $A_o$, $D_o$ are those of the outgoing ray of the same muon at another detector array, as detected as a coincidence event. A tuple of scattering point and a scattering angle: $S_i = (P_i, \Phi_i)$ is determined by the basic *POCA* algorithm ($P_i$ is the point-of-closest-approach or POCA) for each $i$–th event, developed before [3]. The following *MTclear* algorithm [Fig.2] determines whether each voxel $v_m$ ($m \leq M$) in $V$ is either of (1) *threat* type, or (2) *cleared* or non-threat type, or (3) *insufficient-information* type.

*Input: Set of coincidence events E: each $(A_i, D_i, A_o, D_o)$; Voxels $v_m$ ($m \leq M$) in probed volume V; Threshold parameters $\alpha$, c, t*

*Output: Each voxel $v_m$ classified as threat / cleared / insufficient-information types*

// Ray tracing part of the algorithm
1. For each event in E do
2.    Draw lines $l_i$ using $A_i$, $D_i$, and $l_o$ using $A_o$, $D_o$ ;
3.    Find POCA point and angle of scattering between $l_i$ and $l_o$ respectively as $(p, \Phi)$;
4.    If $\Phi$ > threshold angle $\alpha$
5.      Increment POCA count $C_m$ of voxel $v_m$ that contains point p;
   Else
6.      Ray-trace R between detector points $D_i$ to $D_o$ ;
7.      For each voxel $v_m$ on the ray path of R
        Increment straight-track count $T_m$ of voxel $v_m$ ;
     End For loop; // over voxel-wise counting
   End For loop; // on events

// Decision making part of the algorithm
8. For each voxel $v_m$ in V
9.    If $C_m$ > c  then $v_m$ is "threat-type"
10.   Else If $T_m$ > t  then $v_m$ is "cleared"
11.   Else $v_m$ is "insufficient-data"
   End For loop; // over voxels
12. Return all voxels' status in V

*Figure 2. Sketch of the algorithm MTclear for determining the status of voxels in a probed volume V.*

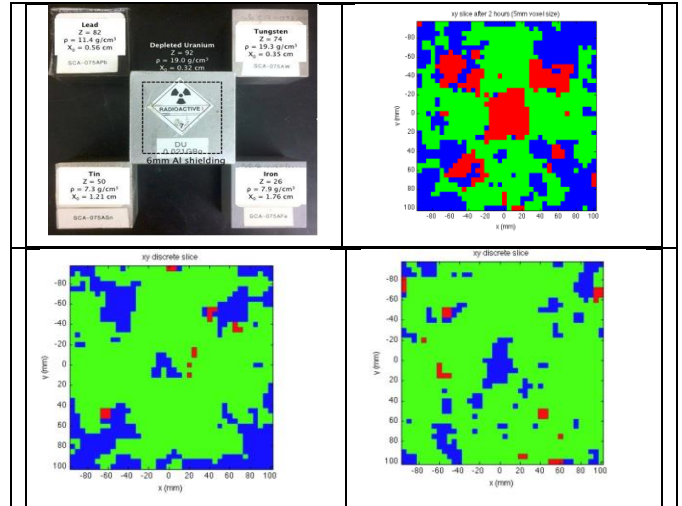

*Figure 3. Upper left is a picture of the configuration of test blocks in the MTS; upper right is a horizontal slice of reconstructed 3D image, where red indicates threat voxels, green indicates cleared voxels, and blue indicates insufficient-data voxels. Bottom left and right show horizontal slices above and below with 2 cm vertical clearance from the physical location of the blocks.*

The status of a voxel $v_m$ may change with more exposure time as more muon tracks are processed. Thus, *MTclear* is potentially a real-time algorithm that may help a human operator or a statistical-decision making automated algorithm to provide a judgment about the threat-possibility in some *V*.

Note that a voxel $v_m$ may have both type of counts, $C_m$ for POCA points, and $T_m$ for straight tracks. However, a decision of "threat" status gets priority over the "cleared" status (steps 9 and 10) in the classification process. Steps 8-12 may be replaced by 3D tomographic visualization of *V*, with color codes for each type of status of a voxel, e.g. red for "threat," green for "cleared," and blue for "insufficient-data," where the color intensity of each of the first two types may indicate the actual respective counts [Fig. 3]. Such a 3D visualization will be helpful for an operator, in the absence of any automated decision making algorithm.

## 3. Results from the MT station

We have experimented by placing five objects in a horizontal plane in the middle of the probed volume *V* in our MTS. The five different blocks of materials used are: lead (Z=62, Northwest corner in the configuration, Fig. 3), tungsten (Z=74, Northeast corner), tin (Z=50, Southwest corner), iron (Z=26, Southeast corner), and DU with 6 mm of aluminum shielding (Z=92, center). Each cubic block is approximately of $5 \times 5 \times 5$ cm$^3$ volume. Fig. 3 shows the configuration in the upper left, and in the upper right corner it shows a slice of the 3D reconstructed image through the horizontal plane of the blocks (color coded as described before) from about 2 hours of collected data (approximately 12,000 tracks).

The bottom left and bottom right of Fig. 3 show a horizontal slice above and one below the blocks, respectively, with about 2 cm clearance from the blocks. We chose $5 \times 5 \times 5$ mm$^3$ voxel dimensions ($v_m$), and the threshold parameters by repeated experimentations.

The *MTclear* algorithm has complexity of the order of $O(max\{kN, M\})$, for *N* number of events, *M* number of voxels, and *k* as the maximum number of voxels on a track over all tracks. The POCA algorithm in steps 1-5 takes $O(N)$ time; the For loop in step 2 runs *N* time, but the ray tracing steps 6-7 may update count-values of *k* voxels for each track or event – for a total order of $O(kN)$ time. The last loop in steps 8-11 (or, the visualization steps) runs over all *M* voxels. On a typical laptop the whole algorithm takes a few minutes to process 2 hours of data from our station.

## 5. Observations and Conclusion

In this work with the *MTclear* algorithm we made two interesting observations. (1) We can clear more voxels faster than we can find sufficient number of scattering (POCA) points within threat voxels for the purpose of quantitatively imaging threat objects within a probed volume *V*. (2) Suspicious threat voxels do get indicated early enough (subject to the choice of the threshold parameter *c*). These two observations together mean that an operator will be able to suspect a probed volume *V* quickly enough (possible in tens of seconds, when "red" threat voxels start appearing in close proximity in space) and then will be able to decide to scan further for a longer time in order to obtain a clear tomographic 3D image of the threat material(s) and its type. In case there is no threat material present in *V,* an operator needs to carefully observe how the voxels or regions of *V* are getting *cleared*, and decide at what time to stop scanning (possibly based on his/her training).

## 6. Future Work

There are many directions that *MTclear* can be extended into. Firstly, an automated decision making process regarding when to stop the scanning could be based on statistics and machine learning. Secondly, we intend to parallelize the ray tracing portion of the algorithm that is ideal for a Graphic Processing Unit (GPU). Thirdly, as the events are generated from the detector arrays, *MTclear* may function as a real-time system by continuously updating the 3D reconstructed image on the operator's monitor. Finally, we may develop a density-based clustering algorithm that does not use voxels as spatial basis, as done in the *PoClust* algorithm [3] for clustering POCA points. An advantage of non-voxelized algorithm is that the forced trade-off between sensitivity versus accuracy of reconstruction is avoided. In voxelized algorithms, sensitivity decreases with voxel size (less data per voxel) but the resolution and statistical accuracy increase when a voxel has sufficient data.